\documentclass[a4paper,10pt]{scrartcl}

\usepackage{amsmath}
\usepackage{amsfonts}
\usepackage{amssymb}
\usepackage{graphicx}
\usepackage[english]{babel}
\usepackage{wasysym}
\usepackage{units}

\newcommand{\p}{\partial}
\newcommand{\reff}[1]{(\ref{#1})}
\newcommand{\vs}[1]{\vspace{#1mm}}
\newcommand{\vsO}{\vspace{.1cm}\hfill\\}
\newcommand{\vsT}{\vspace{.2cm}\hfill\\}

\title{Gravitational Stability and Bulk Cosmology}

\author{Nakia Carlevaro$^{\;a,\,b}$ and Giovanni Montani$^{\;b,\,c,\,d,\,e}$
\vsT\vsO
\emph{\footnotesize $^a$Department of Physics, Polo Scientifico -- Universit\`a degli Studi di Firenze,}\vs{-2.5}\\
\emph{\footnotesize INFN -- Section of Florence, Via G. Sansone, 1 (50019), Sesto Fiorentino (FI), Italy}\\
\emph{\footnotesize $^b$ICRA -- International Center for Relativistic Astrophysics,}\vs{-2.5}\\
\emph{\footnotesize c/o Dep. of Physics - ``Sapienza'' Universit\`a di Roma}\\
\emph{\footnotesize $^c$ Department of Physics - ``Sapienza'' Universit\`a di Roma, Piazza A. Moro, 5 (00185), Rome, Italy}\\
\emph{\footnotesize $^d$ENEA -- C.R. Frascati (Department F.P.N.), Via Enrico Fermi, 45 (00044), Frascati (Rome), Italy}\\
\emph{\footnotesize $^{e}$ ICRANet -- C. C. Pescara, Piazzale della Repubblica, 10 (65100), Pescara, Italy}\\\vsO
{\footnotesize\ttfamily nakia.carlevaro@icra.it\quad\quad\quad montani@icra.it}
}
\date{}
\begin{document}
\maketitle

%
\begin{abstract} \textbf{Abstract:}
We present a discussion of the effects induced by \emph{bulk viscosity} either on the very early Universe stability and on the dynamics associated to the extreme gravitational collapse of a gas cloud. In both cases the viscosity coefficient is related to the energy density $\rho$ via a power-law of the form $\zeta=\zeta_0 \rho^s$ (where $\zeta_0,\,s=const.$) and the behavior of the \emph{density contrast} in analyzed. 

In the first case, matter filling the isotropic and homogeneous background is described by an ultra-relativistic equation of state. The analytic expression of the density contrast shows that its growth is suppressed forward in time as soon as $\zeta_0$ overcomes a critical value. On the other hand, in such a regime, the asymptotic approach to the initial singularity admits an unstable collapsing picture.

In the second case, we investigate the \emph{top-down} fragmentation process of an uniform and spherically symmetric gas cloud within the framework of a Newtonian approach, including the negative pressure contribution associated to the bulk viscous phenomenology. In the extreme regime toward the singularity, we show that the density contrast associated to an adiabatic-like behavior of the gas (which is identified by a particular range of the politropic index) acquire, for sufficiently large viscous contributions, a vanishing behavior which prevents the formation of sub-structures. Such a feature is not present in the isothermal-like collapse. We also emphasize that in the adiabatic-like case bulk viscosity is also responsible for the appearance of a threshold scale (equivalent to a Jeans length) beyond which perturbations begin to increase.

\vsO \emph{PACS}: 98.80.-k, 95.30.Wi, 51.20.+d
\end{abstract}

\section*{Introduction}
Both the extreme regime of a gravitational collapse and the very early stages of the Universe evolution are characterized by a thermodynamics which can not be regarded as settled down into the equilibrium. At sufficiently high temperatures, cross sections of the micro-physical processes are no longer able to restore the thermodynamical equilibrium. Thus we meet stages where the expansion and collapse induce non-equilibrium phenomena. The average effect of having such kind of micro-physics results into dissipative processes appropriately described by the presence of \emph{bulk viscosity} $\zeta$. As shown in \cite{bk76,bk77,bnk79} this kind of viscosity can be phenomenologically described by a function of the energy density $\rho$ as $\zeta=\zeta_0 \rho^s$ where $\zeta_0,\;s=const$.

In the first part of this work we investigate the effects that bulk viscosity has on the stability of the isotropic Universe \cite{nkm051}, \emph{i.e.} the dynamics of cosmological perturbations is analyzed when viscous phenomena affect the zeroth- and first-order evolution of the system. We consider a background corresponding to a Friedmann-Robertson-Walker model filled with ultra-relativistic viscous matter, whose coefficient $\zeta$ corresponds to the choice $s=\nicefrac{1}{2}$ and then we develop a perturbation theory which generalizes the original analysis performed by Lifshitz \cite{l46,lk63}. 

As issue of our first analysis we find that two different dynamical regimes appear when viscosity is taken into account and the transition from one regime to the other one takes place when the parameter $\zeta_0$ overcomes a given threshold value. However in both these stages of evolution the Universe results to be stable as it expands; the effect of increasing viscosity is that the density contrast begins to decrease with increasing time when $\zeta_0$ is over the threshold. It follows that a real new feature arises with respect to the non-viscous ($\zeta_0=0$) analysis when the collapsing point of view is addressed. In fact, if $\zeta_0$ overcomes its critical value, the density contrast explodes asymptotically and the isotropic Universe results unstable approaching the initial singularity. Since a reliable estimation fixes the appearance of thermal bath into the equilibrium below temperatures $\mathcal{O}(10^{16}GeV)$ and this limit corresponds to the pre-inflationary age, our result supports the idea that an isotropic Universe outcomes only after a vacuum phase transition settled down. 

In the second part of this work, we study one of the most attractive challenge of relativistic cosmology concerning the research a self-consistent picture for processes of gravitational instability which connect the high isotropy of the cosmic microwaves background radiation with the striking clumpiness of the actual Universe. An interesting framework is provided by the \emph{top-down} scheme of structure fragmentation which is based on the idea that perturbation scales, contained within a collapsing gas cloud, start to collapse (forming sub-structures) because their mass overcomes the decreasing Jeans value \cite{j902,j28,b56,b57} of the background system. In a work by C. Hunter \cite{h62}, a specific model for a gas cloud fragmentation was addressed and the behavior of sub-scales density perturbations, outcoming in the extreme collapse, was analytically described.

In this respect, we investigate how the above picture is modified by including, in the gas cloud dynamics, the presence of bulk viscosity effects \cite{nkm052}. We start describing the Lagrangian zeroth-order evolution by taking into account the force acting on the collapsing shell as a result of the negative pressure connected to the presence of bulk viscosity. We construct such an extension requiring that the asymptotic dynamics of the collapsing cloud is not qualitatively affected by the presence of viscosity: in this respect we can assume the viscous exponent as $s=\nicefrac{5}{6}$. Then we face the Eulerian motion of the inhomogeneous perturbations living within the cloud. The resulting dynamics is treated in the asymptotic limit to the singularity.

As a result, we show that the density contrast behaves, in the isothermal-like collapse (\emph{i.e.} in correspondence of a politropic index $1\leqslant\gamma<\nicefrac{4}{3}$), as in the non-viscous case $\zeta_0=0$. On the other hand, the damping of perturbations increases monotonically as $\gamma$ runs from $\nicefrac{4}{3}$ to $\nicefrac{5}{3}$; in fact, for such adiabatic-like case, we see that density contrasts asymptotically vanish and no fragmentation processes take place within the cloud when the viscous corrections are sufficiently large. In particular, we observe the appearance of a threshold value for the scale of the collapsing perturbations depending on the values taken by the parameters $\zeta_0$ and $\gamma\in(\nicefrac{4}{3},\nicefrac{5}{3}]$; such a viscous effect corresponds to deal with an analogous of the Jeans length, above which perturbations are able to collapse. However such a threshold value does not ensure the diverging behavior of density contrasts which takes place, in turn, only when a second (greater) critical length is overcome. According to our analysis, if the viscous effects are sufficiently intense, the final system configuration is not a fragmented cloud as a cluster of sub-structures but simply a single object (a black hole, in the present case, because pressure forces are taken negligible).

\section{Bulk Viscosity Effect on the Early Universe}
\subsection{General Remarks}

In order to describe the temporal evolution of the energy density small fluctuation, we develop a perturbations theory on the Einstein equations. We limit our work to the study of space regions having small dimensions compared with the scale factor of the Universe $a$ \cite{ll}. According to this approximation, we can consider a 3-dimensional Euclidean (time dependent) metric as spatial component of the background line element
\begin{equation}
\label{metric}
ds^2=dt^2 - a^2\,(dx^2+dy^2+dz^2)\;.
\end{equation}
In linear approximation, perturbed Einstein equations write as
\begin{equation}
\label{eqeintpert}
\delta R_\mu^\nu - \frac{1}{2}\delta_\mu^\nu \delta R= 8\pi G \delta T_\mu^\nu\;,
\end{equation}
where the term $\delta T_\mu^\nu$ represents the perturbation of the energy-momentum tensor which describes the properties of the matter involved in the cosmological collapse. The perturbations of the Ricci tensor $\delta R_\mu^\nu$ can be written in terms of metric perturbations $h_\mu^\nu=-\delta g_\mu^\nu$, starting from the general expression for the perturbed curvature tensor. For convenience let us now introduce a new temporal variable $\eta$, set by the relation $dt=a\,d\eta$, and use the symbol $(')$ for its derivatives; we moreover impose, without loss of generality, that the synchronous reference system is still preserved in the perturbations scheme: $h_{00}=h_{0\alpha}=0$.

With the above assumptions, the perturbations of the Ricci tensor and the curvature scalar read:
\begin{subequations}
\label{ricci-pert}
\begin{align}
&\delta R_{0}^{0}=-\frac{1}{2a^2}\;h''-\frac{a'}{2a^3}\;h'\;,\\
&\delta R_{0}^{\alpha}=\frac{1}{2a^2}\;(h^{,\,\alpha\,'}-h_{\beta}^{\alpha,\,\beta\,'})\,,\\
&\delta R_{\alpha}^{\beta}= -\frac{1}{2a^2}\;(h^{\gamma,\,\beta}_{\alpha,\,\gamma} +h^{\beta,\,\gamma}_{\gamma,\,\alpha}-h^{\beta,\,\gamma}_{\alpha,\,\gamma} -h^{,\,\beta}_{,\,\alpha})+\nonumber\\
&\qquad\quad-\frac{1}{2a^2}\;h_{\alpha}^{\beta\,''}-\frac{a'}{a^3}\;h_{\alpha}^{\beta\,'}- \frac{a'}{2a^3}\;h'\,\delta_{\alpha}^{\beta}\;\,,\\
&\delta R \,\,\,= -\frac{1}{a^2}\;(h^{\gamma,\,\alpha}_{\alpha,\,\gamma} -h^{,\,\gamma}_{,\,\gamma}) - \frac{1}{a^2}\;h'' - \frac{3 a'}{a^3}\;h'\;.
\end{align}
\end{subequations}

By using these expressions, we are able to rewrite the left-hand side of Einstein eqs. \reff{eqeintpert} through the metric perturbations $h^{\alpha}_{\beta}$.

\paragraph{\textbf{Dynamical Representation of Perturbations}}

Since we use an Euclidean background metric \reff{metric}, we can expand the perturbations in plane waves of the form $e^{i\,\textbf{q}\cdot\textbf{r}}$, where $\textbf{q}$ is the dimensionless comoving wave vector being the physical one $\textbf{k}=\textbf{q}/a$. Here we investigate the gravitational stability properly described by the behavior of the energy density perturbations expressible only by a scalar function; in this sense we have to choose a scalar representation of the metric perturbations \cite{lk63,ll}. Such a picture is made by the scalar harmonics $Q=e^{i\textbf{q}\cdot\textbf{r}}$, from which the following tensor
\begin{equation}
\label{scalar}
Q^{\beta}_{\alpha}=\tfrac{1}{3}\delta^\beta_\alpha Q\;\,,\qquad
P^{\beta}_{\alpha}=[\tfrac{1}{3}\delta^\beta_\alpha-\tfrac{q_\alpha q^\beta}{q^2}]Q\;,
\end{equation}
can be constructed. We can now express the time dependence of the gravitational perturbations through two functions $\lambda(\eta)$, $\mu(\eta)$ and write the tensor $h_\alpha^\beta$ in the form
\begin{equation}
\label{h-form}
h_\alpha^\beta=\lambda(\eta)P_\alpha^\beta+\mu(\eta)Q_\alpha^\beta\;\,,\qquad h=\mu(\eta)Q\;.
\end{equation}

\subsection{Introduction to Viscous Cosmology}

The presence of dissipative processes within the Universe dynamics, as it is expected at temperatures above $\mathcal{O}(10^{16} GeV)$, can be expressed by an additional term in the standard ideal fluid energy-momentum tensor used in the original works by Lifshitz and Khalatnikov \cite{l46,lk63}. Using the conservation law $T^{\nu}_{\mu;\,\nu}=0$ we can express the viscous energy-momentum tensor as
\begin{equation}
\label{T-visc}
T_{\mu\nu}=(\tilde{p}+\rho)u_{\mu}u_{\nu}-\tilde{p}\,g_{\mu\nu}\;,\qquad\;
\tilde{p}= p - \zeta\,u^{\rho}_{;\,\rho}\;,
\end{equation}
where $p$ denotes the usual thermostatic pressure and $\zeta$ is the bulk viscosity coefficient. In the homogeneous models this quantity depends only on time, and therefore we may consider it as a function of the Universe energy density $\rho$. According to literature developments \cite{bk76,bk77,bnk79,m95} we assume that $\zeta$ depends on $\rho$ via a power-law of the form
\begin{equation}
\label{bulk}
\zeta=\zeta_0\,\rho^s\;,	
\end{equation}
where $\zeta_0\geqslant 0$ is a const. and $s$ is a dimensionless parameter. Furthermore, in a comoving system the 4-velocity can be expressed as $u^{0}=\nicefrac{1}{a},\;u^{\alpha}=0$ and the viscous pressure $\tilde{p}$ assumes the form
\begin{equation}
\label{p-prime}	
\tilde{p}= p - 3\,\zeta_0\,\rho^{s}\,\nicefrac{a'}{a^{2}}\;.
\end{equation}
 
Let us now consider the earlier stages of a flat Universe corresponding to $\eta\ll1$ and with the ultra-relativistic equation of state $p=\rho/3$. The Universe zeroth-order dynamics is described by the energy conservation equation and the Friedmann one, which are respectively
\begin{equation}
\label{continuity-friedmann}
\rho'+3\,\frac{a'}{a}\,(\,\rho+\tilde{p}\,)=0\,,\qquad
\frac{a'}{a^{2}}\,=\,\sqrt{\nicefrac{8}{3}\,\pi G\rho}\;.
\end{equation}
In this analysis we assume $s=\nicefrac{1}{2}$ in order to deal with the maximum effect that bulk viscosity can have without dominating the Universe dynamics since it corresponds to a phenomenological issue of perturbations to the thermodynamical equilibrium \cite{b87,b88}. The solutions of the zeroth-order dynamics, for $\,s=\nicefrac{1}{2}$, assume the form  
\begin{equation}
\label{omega}
\rho=C a^{-(2+2\omega)}\;,\;\,a=a_1\,\eta^{1/\omega}\;,\;\,
\omega=1-\chi\,\zeta_0\,,
\end{equation}
being $C$ an integration constant, $\chi=\sqrt{54\pi G}$ and the parameter $a_1=(8\omega^2\pi C G/3)^{1/2\omega}$. Since we consider an expanding Universe, the factor $a$ must increase with positive power of the temporal variable (\emph{i.e.} $\omega>0$) thus we obtain the constraint $0\leqslant\zeta_0<\nicefrac{1}{\chi}$ which ensures this feature.

\subsection{Perturbation Theory in the Early Universe}

Let us now perturb the viscous energy-momentum tensor. Using the synchronous character of the perturbed metric we get the following expressions
\begin{align}
\label{T-pert-visc}
\delta T_{0}^{0}&=\delta \rho\;,\quad
\delta T_{0}^{\alpha}=a\,(\tilde{p}+\rho)\,\delta u^{\alpha}\;,\\
\delta T_{\alpha}^{\beta}&=\delta_{\alpha}^{\beta}\left[-\Sigma^2 \delta \rho + \zeta\left( \delta u^\gamma_{\,,\gamma}+h'/2a^2\right)\right]\;,
\end{align}
here $\Sigma^2\equiv v_s^2 -3 \zeta_0\,s\rho^{s-1}\,a'/a^2\,$ and $v_s$ is the sound speed of the fluid given by $v_s^2=\delta p/\delta\rho$.

The presence of viscosity does not influence the expression of the Ricci tensor and its perturbations, thus we can still keep expressions (\ref{ricci-pert}) and use the perturbed form of the energy-momentum tensor to build up the equations which describe the dynamics of $h_\alpha^\beta$ and $\delta\rho$. It is convenient to choose, as final equations, the ones obtained from the Einstein ones for $\alpha\neq\beta$ and for contraction over $\alpha$ and $\beta$, which read respectively  
\begin{equation}
\label{h-1-visc}
\big(h^{\gamma,\,\beta}_{\alpha,\,\gamma} +h^{\beta,\,\gamma}_{\gamma,\,\alpha} -h^{\beta,\,\gamma}_{\alpha,\,\gamma}-h^{,\,\beta}_{,\,\alpha}\big)+ h^{\beta\,''}_{\alpha}\,+\,\tfrac{{2a'}}{a}\,h^{\beta\,'}_{\alpha}=0\,,
\end{equation}
\begin{equation}
\label{h-2-visc}
\begin{aligned}
\tfrac{1}{2}\,\big(h^{\gamma,\,\alpha}_{\alpha,\,\gamma}&-h^{,\,\gamma}_{,\,\gamma}\big)\big(1+3\Sigma^2\big)+h''+\\
&+\frac{a'}{a}\big(2+3\Sigma^2-12\pi G\frac{a}{a'}\,\zeta\big)\,h'\,+\\ &\quad\quad-\frac{3\zeta}{2a(\tilde{p}+\rho)}\big(\,h^{,\,\alpha\,'}_{,\,\alpha}-\,h^{\gamma,\,\alpha\,'}_{\alpha,\,\gamma}\big)=0\;.
\end{aligned} 
\end{equation}
Furthermore, taking the 00-components of gravitational eqs., we get the expression of the density perturbations
\begin{equation}\label{deltarho}
\delta\rho=\frac{1}{16 \pi G a^2}
(h^{\gamma,\,\alpha}_{\alpha,\,\gamma}- 
h^{,\,\alpha}_{,\,\alpha}+\tfrac{2a'}{a}\,h')\;.
\end{equation}

Substituting in eqs. \reff{h-1-visc},\reff{h-2-visc} the zeroth-order solutions \reff{omega} and the scalar representation of the metric perturbations \reff{h-form}, we can get, respectively, two equations for $\lambda$, $\mu$ which read
\begin{equation}
\label{eq-la-visc}
\lambda''+\tfrac{2}{\omega\eta}\,\lambda'-\tfrac{q^2}{3}\,\left(\lambda+\mu\right)=0\;,
\end{equation}
\begin{equation}
\label{eq-mu-visc}
\begin{aligned}
&\mu''+\Big(\frac{2+3\Sigma^2}{\omega\eta}\Big)\mu'-\Big(\frac{12 \pi \sqrt{C}G\zeta_0} {a_1^{1+\omega}\,\eta^{1+1/\omega}}\Big)\mu'+\\
&+\tfrac{q^2}{3}\left(\lambda+\mu\right)
\left(1+3\Sigma^2\right)+\frac{q^2\zeta_0\eta\,\left(\mu'+\lambda'\right)}
{\nicefrac{4\sqrt{C}}{3a_1^\omega}-\nicefrac{3\zeta_0}{\omega}}=0.
\end{aligned} 
\end{equation}

It is worth nothing that, among the solutions, there are some which can be removed by a simple transformation of the reference system, compatible with its synchronous character. They do not represent any real physical change in the metric and we must exclude them from the dynamics. Such fictitious solutions, in the ultra-relativistic limit, assume the form $\;\lambda-\mu=const.\;$ and $\;\lambda+\mu\sim1/\eta^2$ \cite{ll}.

\subsection{Behavior of the Density Contrast}

Let us now study the gravitational collapse dynamics of the primordial Universe near the initial \emph{Big-Bang} in the limit $\eta\ll1$. As in Lifshitz work \cite{lk63}, we analyze the case of perturbations scale sufficiently large to use the approximation $\eta q\ll1$. In our scheme eqs. (\ref{eq-la-visc}) and (\ref{eq-mu-visc}) admit asymptotic analytic solutions for the functions $\lambda$ and $\mu$; in the leading order $\lambda$ takes the form
\begin{equation}
\label{sol-l-visc}
\lambda=\frac{C_1}{\eta^{2/\omega-1}}\,+\,C_2\;,
\end{equation}
where $C_1$, $C_2$ are two integration constants. Substituting this expression in eq. (\ref{eq-mu-visc}) we get, in the same order of approximation, the behavior of the function $\mu$ as
\begin{equation}
\label{sol-mu-visc}
\mu=\frac{\tilde{C}_1}{\eta^{1/\omega-3}}\,+\,C_2\;,
\end{equation}
where we have excluded the non-physical solutions introduced above. The constant $\tilde{C}_1$ is given by the expression $\tilde{C}_1=\nicefrac{A}{B}(3-\nicefrac{1}{\omega})$, $A$ and $B$ being constants having the form
{\footnotesize\begin{equation}\nonumber
\label{A-B}
A=\frac{C_1\,q^2}{3}\left(1+3\Sigma^2\right)+\frac{C_1(1-2/\omega)q^2\zeta_0}{4\sqrt{C}/3a_1^\omega-3\zeta_0/\omega}\,,\; B=\frac{12 \pi \sqrt{C} G\zeta_0}{a_1^{1+\omega}}\;.
\end{equation}}

Let us now write the final form of perturbations pointing out their temporal dependence in the viscous Universe. The metric perturbations (\ref{h-form}) become
\begin{equation}
\label{h-visc}
h^{\beta}_{\alpha}=\frac{C_1}{\eta^{2/\omega-1}}\,P^{\beta}_{\alpha}+ \frac{\tilde{C}_1}{\eta^{1/\omega-3}}\,Q^{\beta}_{\alpha}+C_2\left(Q^{\beta}_{\alpha}+P^{\beta}_{\alpha}\right)\;,
\end{equation}
and, by \reff{deltarho} and \reff{omega}, the density contrast $\delta=\nicefrac{\delta\rho}{\rho}$ reads
\begin{equation}
\label{contr-visc}
\delta=F[C_1\eta^{3-\nicefrac{2}{\omega}}+C_2\eta^2+C_3\eta^{3-\nicefrac{1}{\omega}}
+\tilde{C}_1\eta^{5-\nicefrac{1}{\omega}}],
\end{equation}
where $C_3=3A/q^2\omega B$, and $F=\omega^2Qq^{2}/9$. 

We have now to impose the conditions expressing the smallness of perturbations at a given initial time $\eta_0$. The inequalities $h^{\beta}_{\alpha}\ll1$ and $\delta\ll1$  yield only two fundamental constraints for the integration constants: 
\begin{equation}
C_1\ll\eta_0^{2/\omega-1}\;,\qquad C_2\ll1\;,
\end{equation}
for any $\omega$-value within the interval $(0,\,1]$. Furthermore we find an additional condition which involves the wave number $q$ and the integration constant $C$; in particular a rough estimation for $\omega<1/3$ of the inequalities $\tilde{C}_1\ll\eta_0^{1/\omega-3}$ and $C_3\ll\eta_0^{1/\omega-3}$ yields the condition $q\ll(GC\eta_0)^{-1/2\omega}$ which ensures the smallness of the cosmological perturbations.
 
Using the hypothesis $\eta\ll1$ we can get the asymptotic form of the corrections to the cosmological background. The exponents of the variable $\eta$ can be positive or negative according to the value of the viscous parameter $\omega(\zeta_0)$. This behavior produces two different regimes of the density contrast evolution: 

\paragraph{\textbf{Case} $\;\;0\leqslant\zeta_0<\nicefrac{1}{3\chi}\;\,\,$}
Here perturbations increase forward in time. This behavior corresponds qualitatively to the same picture of the non-viscous Universe (obtained setting $\zeta_0=0$) in which the expansion can not, nevertheless, imply the gravitational instability: if we consider the magnitude order $\eta\sim1/q$, the constraints on $C_1$, $C_2$ imply that $\delta$ remains small even in the higher order of approximation. This behavior yields the gravitational stability of the primordial Universe.

\paragraph{\textbf{Case} $\;\;\nicefrac{1}{3\chi}<\zeta_0<\nicefrac{1}{\chi}\;\,\,$} 
In this regime the density contrast is suppressed behaving like a negative power of $\eta$. When the density contrast results to be increasing, the presence of viscosity induces a \emph{damping} of the perturbation evolution in the direction of the expanding Universe, so the cosmological stability is fortified since the leading $\eta$ powers are smaller than the non-viscous ones obtained setting $\zeta_0=0$. 

In this case the density fluctuation decreases forward in time but the most interesting result is the instability which the isotropic and homogeneous Universe acquires in the direction of the collapse toward the Big-Bang. For $\zeta_0>\nicefrac{1}{3\chi}$ the density contrast diverges approaching the cosmological singularity, \emph{i.e.} for $\eta\rightarrow0$. In this regime, scalar perturbations destroy asymptotically the primordial Universe symmetry. The dynamical implication of this issue is that an isotropic and homogeneous stage of the Universe can not be generated, from generic initial conditions, as far as the viscosity becomes smaller than the critical value $\zeta_0^{*}=\nicefrac{1}{3\chi}$.


\section {Spherically Symmetric Gas Cloud Fragmentation}\label{Sec.2}

In this second part, we present an hydrodynamical analysis of a spherically symmetric gas cloud collapse. This model was firstly proposed in a work by C. Hunter \cite{h62}, where he supposed that a perfect fluid cloud becomes unstable with respect to its own gravitation and begin to condense. The collapsing cloud is assumed to be the dynamical background on which studying, in a Newtonian regime, the evolution of density perturbations generated on this basic flow. Such an analysis is suitable for the investigation of the cosmological structures formation in the \emph{top-down} scheme \cite{be80,bs83} since it deals with the sub-structures temporal evolution compared with the basic flow of the gravitational collapse.

\subsection{Unperturbed dynamics}

The unperturbed flow is supposed to be homogeneous, spherically symmetric and initially at rest. Furthermore the gravitational forces are assumed to be very much greater than the pressure ones, which are therefore neglected in the zeroth-order analysis. In such an approach the gas results to be unstable, since there are no forces which can contrast the collapse, and the condensation starts immediately.

The basic flow is governed by the Lagrangian motion equation of a spherically symmetric gas distribution which collapses under the only gravitational action. As in the previous approach (see eqs. \reff{T-visc}, \reff{bulk}), in order to include bulk viscosity effects in the dynamics, we introduce the bulk pressure
\begin{equation}
\tilde{p}=p-\zeta_0\rho^s\,u^{\mu}_{;\,\mu}\;,
\end{equation}
where $u_\mu=(1,\textbf{0})$ is the shell comoving 4-velocity. In the Newtonian limit we consider, the metric can be assumed as a flat Minkowskian one expressed in the usual spherical coordinates and the metric determinant $g$ becomes $g=-r^{4}sin^{2}\theta$. In this case, for the 4-divergence $u^{\mu}_{;\,\mu}$ we immediately obtain: $u^{\mu}_{;\,\mu}=2\dot{r}/r\;$. 

Considering the basic flow density as $\rho=M/(\nicefrac{4}{3}\pi r^{3})$ and the pressure force acting on the collapsing shell of the form $F_{\tilde{p}}=\tilde{p}\,4\pi r^{2}$, the Lagrangian motion equation for a viscous fluid becomes now
\begin{equation}
\label{eq-lagr}
\frac{\partial^2r}{\partial t^2}=-\frac{GM}{r^2}-\frac{C}{r^{3s-1}}\,\frac{\p r}{\p t}\;,
\end{equation}
where $C=8\pi\zeta_0\left(3M/4\pi\right)^{s}$. The origin $O$ is taken at the center of the gas, $r$ is the radial distance, $G$ the gravitational constant and $M$ the mass of the gas inside a sphere of radius $r$. In what follows, we shall suppose that the gas was at a distance $a$ from $O$ in correspondence to the initial instant $t_0$; this distance $a$ identifies a fluid particle and will be used as a Lagrangian independent variable so $r=r(a,t)$ and the following parametrization can be introduced 
\begin{equation}
\label{r-a}
r=a\cos^2\beta\;.
\end{equation}
Here $\beta=\beta(t)$ is a time dependent function such that $\beta(t_0)=0$ and $\beta(0)=\pi/2$ since we choose the origin of time to have $t=0$ when $r=0$ and $t_0$ takes negative values. 

Eq. \reff{eq-lagr} must be integrated to obtain the evolution of the radial velocity $v=\nicefrac{\p r}{\p t}$ and the density $\rho$ of the unperturbed flow. Let us now require that the viscosity does not influence the final form of the velocity which in the non-viscous Hunter analysis (\emph{i.e.} $\zeta_0=0$) is proportional to $\nicefrac{1}{\sqrt{r}}$ \cite{h62}. Substituting $v=\nicefrac{B}{\sqrt{r}}$ into eq. \reff{eq-lagr} we see that, in correspondence to the choice $s=\nicefrac{5}{6}$, it is again a solution as soon as $B=C-\sqrt{C^{2}+2GM}\;,$ where $B$ assumes only negative values. Although this dynamics is analytically integrable only for the particular value $s=\nicefrac{5}{6}$, the obtained behavior $v\sim\nicefrac{1}{\sqrt{r}}$ remains asymptotically (as $r\rightarrow0$) valid if the condition $s<\nicefrac{5}{6}$ is satisfied. 

Using such a solution we are able to build an explicit form of the quantity $\beta$ defined by \reff{r-a}. After standard manipulations we obtain the relation
\begin{equation}
\label{beta-t}
\cos\beta^{3}=3A\,(-t)\;,
\end{equation}
where $A$ is defined to be $A=-B/2a^{3/2}$. 
 
It is more convenient to use an Eulerian representation of the flow field. To this end, once solving the well known Poisson equation for the gravitational potential $\phi$, we obtain the unperturbed solutions describing the background motion; all these quantities take the explicit forms
\begin{subequations}
\label{basic-flow}
\begin{align}
\label{V-basic-flow}
& \textbf{v}=[v,0,0]\;,\quad v=-2r\dot{\beta}\tan\beta\;,\\
\label{rho-basic-flow}
& \rho=\rho_0\cos^{-6}\beta\;,\\
\label{phi-basic-flow}
& \phi=-2\pi\rho_0 G\left(a^2-r^2/3\right)\cos^{-6}\beta\;,
\end{align}
\end{subequations}
where ($\,\dot{}\,$) denotes the derivate with respect to time and the non-radial components of velocity must vanish since we are considering a spherical symmetry.

\subsection{First-order perturbation theory}

The zeroth-order motion of a viscous basic flow which collapses under the action of its own gravitation was discussed above. We shall now suppose that small disturbances appear on this field; the perturbations evolution can be described by the Eulerian equations in presence of dissipative processes which come out from the thermodynamical irreversibility of the collapse process and are due to the microphysics of non-equilibrium \cite{ll-fluid}.

Perturbations are investigated in the Newtonian limit starting from the well known continuity, Euler-Navier-Stokes and Poisson equations. In such a picture, we are now interested to study the effects of the thermostatic pressure on the perturbations evolution. We shall therefore consider terms due to the pressure $p$ in the motion equations, which read 
\begin{subequations}
\label{eq-unpert-0}
\begin{align}
\label{continuity-unpert-0}
&\dot{\rho}+\nabla\cdot(\rho\,\textbf{v})=0\;,\\
\label{eulero-unpert-0}
&\dot{\textbf{v}}+(\textbf{v}\cdot\nabla)\cdot\textbf{v}=
-\nabla\phi-\nicefrac{1}{\rho}\,\nabla
p+\nicefrac{\zeta}{\rho}\,\nabla(\nabla\cdot\textbf{v})\;,\\
\label{poisson-unpert-0}
&\nabla^2\phi=4\pi G\rho\;. 
\end{align}
\end{subequations}
The gas is furthermore assumed to be barotropic, \emph{i.e.} the pressure depends only by the background density by 
\begin{equation}
p=\kappa\rho^{\gamma}\;,
\end{equation}
where $\kappa$, $\gamma$ are constants and $1\leqslant\gamma\leqslant\nicefrac{5}{3}$. By this expression we are able to distinguish a set of different cases related to different values of the politropic index $\gamma$. The asymptotic value $\gamma=1$ represents an isothermal behavior of the gas cloud; the case $\gamma=\nicefrac{5}{3}$ describes, instead, an adiabatic behavior and it will be valid when changes are taking place so fast that no heat is transferred between elements of the gas. We can suppose that intermediate values of $\gamma$ will describe intermediate types of behavior between the isothermal and adiabatic ones.  

In this model, zeroth-order solutions \reff{basic-flow} are already verified since the pressure gradient, in the homogeneity hypothesis, vanishes and the pressure affects only the perturbative dynamics.
  
Let us now investigate first-order fluctuations around the unperturbed solutions, \emph{i.e.} we replace the perturbed quantities: $(\textbf{v}+\delta\textbf{v})$, $(\rho+\delta\rho)$, $(\phi+\delta\phi)$ and $(p+\delta p)$ in eqs. \reff{eq-unpert-0}. Following the line of the Hunter work, equations for the perturbed quantities $\delta\textbf{v}$, $\delta\rho$ and $\delta\phi$ can be built. Skipping many technicality, we are able to manipulate the system of the motion eqs. and write down an unique equation which which governs the evolution of density perturbations $\delta\rho$ in the viscous regime. Taking into account the parameterization \reff{r-a} we get
\begin{align}
\label{eq-fundamental}
&(\cos^{10}\beta\dot{\delta\rho}-6\sin\beta\cos^9\beta\dot{\beta}\delta\rho)^{\dot{}}
-4\pi G\rho_{0}\cos^4\beta\delta\rho=\nonumber\\
&=({v_s}^2\,\cos^6\beta-6\,\nicefrac{\zeta}{\rho_0}\,\sin\beta\cos^{11}\beta\,\dot{\beta})\;D^2\delta\rho+\nonumber\\
&\hspace{3.8cm}+\nicefrac{\zeta}{\rho_0}\,\cos^{12}\beta \;D^{2}\,\dot{\delta\rho}\;.
\end{align}
Here time differentiation is taken at some fixed comoving radial coordinate, $v_s$ is the sound speed $v_s^{2}= \p p/\p\rho$ and $D^2$ is the Laplace operator as written in comoving spherical coordinates.

In order to study the temporal evolution of density perturbations, we assume to expand $\delta\rho$ in plane waves of the form
\begin{equation}
\label{fourier-trasform}
\delta\rho(\textbf{r},t)\rightarrow\delta\rho\,(t)\;e^{-i\,\textbf{k}\cdot\textbf{r}}\;,
\end{equation}
where $1/k$ (with $k=|\textbf{k}|$) represents the initial length scale of the considered fluctuation. According to this assumption, we are able to write the asymptotic form of eq. \reff{eq-fundamental} near the end of the collapse as $(-t)\rightarrow0$. In this case, we can make the approximation $sin\beta\approx1$ and the quantity $cos\beta$ is given by \reff{beta-t}. 

The background motion equations were derived for a particular value of the viscosity parameter $s=\nicefrac{5}{6}$: substituting the basic flow density given by \reff{rho-basic-flow} in the standard expression of the bulk viscosity \reff{bulk}, we obtain
$\zeta\,=\,\zeta_0\rho_0^{\nicefrac{5}{6}}\,\,cos^{-5}\beta\;.$ With these assumptions, eq. \reff{eq-fundamental} now reads
\begin{align}
\label{eq-asym}
&(-t)^2\,\ddot{\delta\varrho}-\bigg[\frac{16}{3}-\frac{\lambda}{3A}\bigg](-t)\,\dot{\delta\varrho}+\nonumber\\
&+\bigg[\frac{14}{3}-\frac{4\pi G \rho_0}{9 A^{2}}+\frac{v_0^2\, k^2(-t)^{8/3-2\gamma}}{(3A)^{2\gamma-2/3}}-\frac{2\lambda}{3A}\bigg]\delta\varrho=0,
\end{align}
where $v_0^{2}=\kappa\gamma\rho_0^{\gamma-1}$ and the viscous parameter $\lambda$ is given by
\begin{equation}
\label{lambda}
\lambda\,=\,\zeta_0\,\rho_0^{-\nicefrac{1}{6}}\,k^{2}\;.
\end{equation}

\subsection{Asymptotic Analysis of The Density Contrast}

A complete solution of eq. \reff{eq-asym} involves Bessel functions of first and second species $J$ and $Y$ respectively and it explicitly reads
\begin{equation}
\label{drho-solution}
\delta\varrho=\,C_1\,G_1(t)\,+\,C_2\,G_2(t)\;,
\end{equation}
where $C_1$, $C_2$ are integration constants and the functions $G_1$ and $G_2$ are defined to be
\begin{align}
\label{g-functions-1}
&G_1(t)=(-t)^{-\frac{13}{6}+\frac{\lambda}{6A}}\;J_n\big[q(-t)^{\nicefrac{4}{3}-\gamma}\big]\;,\\
\label{g-functions-2}
&G_2(t)=(-t)^{-\frac{13}{6}+\frac{\lambda}{6A}}\;Y_n\big[q(-t)^{\nicefrac{4}{3}-\gamma}\big]\;,
\end{align}
having set the Bessel parameters $n$ and $q$ as  
{\small\begin{subequations}
\begin{align}
&n=[A^2-2\lambda A+\lambda^2+16\pi G \rho_0]^{\frac{1}{2}}\;/\,(6A(\nicefrac{4}{3}-\gamma))\;,\nonumber\\
&q=-k v_0(3A)^{1/3-\gamma}\;/\,(4/3-\gamma)\;.\nonumber
\end{align}
\end{subequations}}

We now proceed, in order to study the asymptotic evolution of the solution \reff{drho-solution}, analyzing the cases $1\leqslant\gamma<\nicefrac{4}{3}$ and $\nicefrac{4}{3}<\gamma\leqslant\nicefrac{5}{3}$ separately, since Bessel functions have different limits connected to the magnitude of their argument. In the asymptotic limit to the singularity, the isothermal-like case is characterized by a positive time exponent inside Bessel functions so $q(-t)^{4/3-\gamma}\ll1$, on the other hand, in the adiabatic-like behavior, we obtain $q(-t)^{4/3-\gamma}\gg1$. For an argument mush less than unity Bessel functions $J$ and $Y$ behave like a power-law of the form {\footnotesize{$J_n(x)\sim x^{+n}$}}, {\footnotesize{$Y_n(x)\sim x^{-n}$}}, for $x\ll1$. And in the adiabatic-like case argument becomes much gather than unity and they assume an oscillating behavior like {\footnotesize{$J_n(x)\sim x^{-1/2}cos(x)$}}, {\footnotesize{$Y_n(x)\sim x^{-1/2}sin(x)$}}, for $x\gg1$. 

\paragraph{\textbf{Isothermal-like Case}}

In this first regime, an asymptotic form of functions $G$ can be found as follow
\begin{equation}
\label{g-functions-iso}
G_{1,2}^{ISO}=c_{1}\;\,(-t)^{-\frac{13}{6}+\frac{\lambda}{6A}\pm(\frac{4}{3}-\gamma)n}\end{equation}
where $c_1$ and $c_2$ are constants quantities. The condition which implies the density perturbations collapse is that at least one of $G$ functions diverges as $(-t)\rightarrow0$. An analysis of time exponents yields that $G_1$ diverges if $\lambda<7A-2\pi G\rho_0/3A$ but, on the other hand, $G_2$ is always divergent for all $\lambda$. These results imply that, in the isothermal case, perturbations always condense. 

Let us now compare this collapse with the basic flow one; the background density evolves like $cos^{-6}\beta$ (see \reff{rho-basic-flow}) that is, using \reff{beta-t}
\begin{equation}
\label{rho-t-basic-flow-visc}
\rho\sim(-t)^{-2}\;.
\end{equation}
In the non-dissipative case ($\lambda=0$) perturbations grow more rapidly with respect to the background density involving the fragmentation of the basic flow independently on the value of $\gamma$; in presence of viscosity the density contrast assumes the asymptotic form
\begin{equation}
\label{delta-iso}
\delta^{\,ISO}\sim(-t)^{-\frac{1}{6}+\frac{\lambda}{6A}-\frac{1}{6A}\sqrt{}[A^2-2\lambda A+\lambda^2+16\pi G \rho_0]}\;.
\end{equation}
Here the exponent is always negative and it does not depend on $\gamma$, this implies that $\delta^{\,ISO}$ diverges as the singularity is approached and real sub-structures are formed involving the basic flow fragmentation. This issue means that the viscous forces do not have enough strength to contrast an isothermal perturbations collapse in order to form of an unique structure. 

\paragraph{\textbf{Adiabatic-like Case}}

For $\nicefrac{4}{3}<\gamma\leqslant\nicefrac{5}{3}$, $J$ and $Y$ assume an oscillating behavior. In this regime functions $G$ read
\begin{equation}
\label{g-functions-adb}
G_{1,2}^{\,ADB}=\tilde{c}_{1,2}\;\;^{\cos}_{\sin}\big[q(-t)^{4/3-\gamma}\big]\;(-t)^{\frac{\gamma}{2}-\frac{17}{6}+\frac{\lambda}{6A}}\;,	
\end{equation}
where $\tilde{c}_{1,2}$ are constants. Following the isothermal approach, we shall now analyze the time power-law exponent in order to determine the collapse conditions. $G$ functions diverge, involving perturbations condensation, if the parameter $\lambda$ is less than a threshold value: this condition reads $\lambda<17A-3A\gamma$ (for a given value of the index $\gamma$). Expressing $\lambda$ in function of the wave number \reff{lambda}, we outline, for a fixed viscous parameter $\zeta_0$, a constraint on $k$ which is similar to the condition appearing in the Jeans model \cite{j902}, \cite{j28}. The threshold value for the wave number is given by the relation
\begin{equation}
\label{k-critic}
k_C^{2}=(17A-3\gamma A)\rho_0^{\nicefrac{1}{6}}\,/\,\zeta_0
\end{equation}
and therefore the condition for the density perturbations collapse, \emph{i.e.} $\delta\rho^{ADB}\rightarrow\infty$, reads $k<k_C$, recalling that, in the Jeans model for a static background, the condition for the collapse is $k<k_J=[4\pi G\rho_0\,/\,v_s^{2}\,]^{\nicefrac{1}{2}}$. It is to be remarked that, in absence of viscosity ($\zeta_0=0$), expression \reff{k-critic} diverges implying that all perturbations scales can be conducted to the collapse. On the other hand, if we consider perturbations of fixed wave number, they asymptotically decrease as $(-t)\rightarrow0$ for $\lambda>17A-3A\gamma$. Thus for each $k$ there is a value of the bulk viscosity coefficient over which the dissipative forces contrast the formation of sub-structures.

If $k<k_C$, perturbations oscillate with ever increasing frequency and amplitude. For a non-zero viscosity coefficient, the density contrast evolves like
\begin{equation}
\label{drho-adb}
\delta^{\,ADB}\,\sim\,(-t)^{\frac{\gamma}{2}-\frac{5}{6}+\frac{\lambda}{6A}}\;.
\end{equation}
A study of the time exponent yields a new threshold value. If $\lambda<5A-3A\gamma$, \emph{i.e.} the viscosity is enough small, sub-structures form; on the other hand, when the parameter $\zeta_0$, or the wave number $k$, provides a $\lambda$-term overcoming this value, the perturbations collapse is so much contrasted that no fragmentation process occurs. In other words, if $\lambda>5A-3A\gamma$ we get $\delta^{\,ADB}\rightarrow0$, \emph{i.e.} for a given $\gamma$ there is a viscous coefficient $\zeta_0$ enough large ables to prevents the sub-structures formation. 

It is remarkable that in the pure adiabatic case, $\gamma=\nicefrac{5}{3}$, dissipative processes, of any magnitude order, contrast the fragmentation because, while the Jeans-like length survives, the threshold value for sub-structures formation approaches infinity. We can conclude that, in the case $\nicefrac{4}{3}<\gamma\leqslant\nicefrac{5}{3}$, the fragmentation in the top-down scheme is deeply unfavored by the presence of bulk viscosity which strongly contrasts the density perturbations collapse.

\section{Concluding remarks}
To complete the analysis of the gravitational stability in presence of dissipative affects, we now point out some relevant feature about the validity of our picture concerning both the relativistic analysis and the Newtonian fragmentation:

\paragraph{Shear Viscosity}\quad In our approach, the hypothesis of the Universe isotropy and the fact that the shell results comoving with the collapsing background, imply that there are no displacements between parts of fluid with respect to ones other in both cases. Dissipative processes are therefore related only to thermodynamical properties of the fluid compression and can be phenomenologically described by the presence of bulk viscosity. Furthermore, in this model we are able to neglect the so-called \emph{shear viscosity} since it is connected with processes of relative motion among different parts of the fluid.

\paragraph{Early Universe Dynamics}\quad In the fist Universe analysis, we fix the value $s=\nicefrac{1}{2}$ in order to deal with the maximum effect that bulk viscosity can have without dominating the dynamics. In fact, the notion of this kind of viscosity corresponds to a phenomenological issue of perturbations to the thermodynamical equilibrium. In this sense, we remark that, if $s>\nicefrac{1}{2}$, the dissipative effects become dominant and non-perturbative. Moreover, if we assume the viscous parameter $s<\nicefrac{1}{2}$, the dynamics of the early Universe is characterized by an expansion via a power-law $a(t)\sim t^{2/3\gamma}$ starting from a perfect fluid Friedmann singularity at $t=0$ (here $\gamma$ is identify by the relation $p=(\gamma-1)\,\epsilon$). After this first stage of evolution, where viscosity does not affect at all the dynamics, the Universe inflates in the limit $t\rightarrow\infty$ (\emph{i.e.} out of our approximation scheme) to a viscous deSitter solution characterized by $a(t)\sim e^{H_0t}$, $H_0=\sqrt{\epsilon_0}/3=\nicefrac{1}{3} (\zeta_0\sqrt{3}/\gamma)^{1/(1-2s)}$ \cite{b87,b88}. Since, in this work, we deal with the asymptotic limit $t\rightarrow0$, we only treat the case $s=\nicefrac{1}{2}$ in order to quantitatively include dissipative effects in the primordial dynamics. 

\paragraph{Zeroth-order Cloud Dynamics}\quad We now clarify why the choice $s=\nicefrac{5}{6}$ is appropriate to a consistent treatment of the asymptotic viscous collapse. We start by observing that bulk viscous effects can be treated in a predictive way only if they behave as small corrections to the thermodynamical system. To this respect we have to require that the asymptotic collapse is yet appropriately described by the non viscous background flow. Taking into account eq. \reff{eq-lagr} it is easy to infer that in the asymptotic limit as $r\rightarrow0$, the non viscous behavior $v\sim\nicefrac{1}{\sqrt{r}}$ is preserved only if $s\leqslant\nicefrac{5}{6}$; in fact, in correspondence to this restriction, the viscous correction, behaving like $\mathcal{O}(r^{-3s+1/2})$, is negligible with respect to the leading order $\mathcal{O}(r^{-2})$ when the singularity is approached. On the other hand, it is immediate to verify that, as $(-t)\rightarrow0$, the viscous terms in $\dot{\delta\rho}$ and in $\delta\varrho$ respectively are negligible and the perturbation dynamics \reff{eq-asym} matches asymptotically the non-viscous Hunter result if $s<\nicefrac{5}{6}$. Matching together the above considerations for the zeroth- and first-order respectively, we see that $s=\nicefrac{5}{6}$ is the only physical value which does not affect the background dynamics but makes important the viscous corrections in the asymptotic behavior of the density contrasts.

\paragraph{Validity of the Newtonian Approximation}\quad Since the analysis of the cloud collapse addresses Newtonian dynamics while the cloud approaches the extreme collapse, it is relevant to precise the conditions which ensure the validity of such a scheme. The request that the shell corresponding to the radial coordinate $r$ lives in the Newtonian paradigm leads to impose that it remains greater than its own \emph{Schwarzschild Radius}, $r(t)\gg 2GM(a)\;,$ where $M(a)=\rho_0\,(\textnormal{\tiny{$\frac{4}{3}$}}\pi a^3)$. About the dynamics of a physical perturbations scale $l=(2\pi/k)cos\beta^{2}$ (here $cos\beta^{2}$ plays the same role of a cosmic scale factor), its Newtonian evolution is ensured by the linear behavior, as soon as, the Schwarzschild condition for the background holds. More precisely a perturbations scale is Newtonian if its size is much smaller than the typical space-time curvature length, but for a weak gravitational field this requirement must have no physical relevance. To explicit such a condition, we require that the physical perturbations scale is much greater than its own Schwarzschild Radius, which leads to the inequality $k\gg\chi(-t)^{-1/3}$, where $\chi=\left[\textnormal{\tiny{$\frac{4}{3}$}}(2\pi)^{3}G\rho_0(3A)^{-2/3}\right]^{1/2}$. Such a condition tells us which modes are Newtonian within the shell.


\end{document}